\documentclass[12pt]{article}
\usepackage{amsmath}
\usepackage{amssymb}
\usepackage{graphicx}
\usepackage[cp1251]{inputenc}
\usepackage[english]{babel}
\newcommand{\sign}{\mathop{\mathrm{sign}}}

\title{A two-parametric family of exactly solvable
Dirac Hamiltonians}
\author{Ekaterina Pozdeeva, \\
{\it Semenov Institute of Chemical Physics Russian Academy of Sience,}\\
{\it Moscow, Russia}\\
{\it ekatpozdeeva@mail.ru}}
\date{}
\begin{document}
\selectlanguage{english} \maketitle
\begin{abstract}
We construct a two-parametric family of exactly solvable Dirac
Hamiltonians by the Darboux transformation method. We obtain
intertwining relations between different members of the Hamiltonian
family. We investigate the spectral properties of the obtained
Hamiltonians and the explicit forms of their eigenfunctions.
\end{abstract}
\section{Introduction}
The Darboux transformation method
\cite{Darbu1,Darbu2,Crum,paper1Matveev,paper2Matveev,Salle,Matv}
is one of the basic methods of construction exactly solvable
models of mathematical physics
\cite{Krein,Cooper,Nieto,preprintMatveev,Chang,Pozdeeva1,Pozdeeva2,Yurov,Yurov1,Andrianov}.

The idea of the method has been proposed by G. Darboux in 1882
\cite{Darbu1}. The method has been developed by  M. Crum in 1954
\cite{Crum}.

Modern concept of the Darboux transformation method has been
constructed  by V. B. Matveev. In 1979 V. B. Matveev generalized
and reformulated the Darboux--Crum results for the cases of
infinite hierarchies linear and nonlinear differential equations
in partial derivatives and some their generalizations (for
example, differential--difference and matrix equations), including
nonstationary  nonlinear Schr\"o\-din\-ger, Korteweg--de Vries,
Kadomtsev--Petviashvili equations and others
\cite{paper1Matveev,paper2Matveev}. Numerous realizations of the
Darboux transformation method \cite{paper1Matveev,paper2Matveev}
have been summarized in monograph \cite{Matv}.

It should be noted, that the results of works
\cite{paper1Matveev,Salle,Matv} contain the cases of stationary
and  nonstationary two-component Dirac equation. In  explicit form
the systematic development of the Darboux--Crum transformation
me\-thod for the one--dimensional two-component Dirac equation is
present in \cite{annphys2003v305p151}.

From \cite{Matv} it is evidently that the Darboux transformations
can be applied to the stationary four-component Dirac equation.

In this paper we use the Darboux transformation method for the
four--component Dirac equation.

The structure of the present paper is as follows. In Section 2 we
generate four  one-soliton Dirac Hamiltonians applying the Darboux
transformation method for free Dirac Hamiltonians and consider the
intertwining relations between the obtained Hamiltonians. In Section
3 we construct two-parameter family of exactly solvable multisoliton
solutions. In Section 4 we consider spectrum of the obtained
Hamiltonians and the explicit forms of their eigenfunctions.
\section{One-soliton Hamiltonians}
Let us consider the one--dimensional four-component stationary
Dirac equation
\begin{eqnarray}\label{90}
H_0\psi(x)=E\psi(x),\quad H_0=-i\alpha _1 \partial_x+V_0,\quad\alpha_1=\left(%
\begin{array}{cc}
  0 & \sigma_1 \\
  \sigma_1 & 0 \\
\end{array}%
\right),
\end{eqnarray}
where $\sigma_1=\left(%
\begin{array}{cc}
  0 & 1 \\
  1 & 0 \\
\end{array}%
\right)$ is the Pauli  matrix, $V_0$ is $4\times 4$ Hermitian
matrix, $\psi=\left(
  \psi_1,\psi_2,\psi_3,\psi_4\right)^t$ is four-component spinor.\\
  Let $4\times 4$ matrix $u(x)$ is the solution of the matrix Dirac equation
\begin{eqnarray}\label{90}
H_0u(x)=u(x)\Lambda,
\end{eqnarray}
where $\Lambda$ is any nonsingular matrix.\\
Define the Darboux transformation operator
\begin{eqnarray}
  L=\frac{\partial}{\partial x}-\frac{du(x)}{dx}u^{-1}
\end{eqnarray}
and consider the intertwining relation
\begin{eqnarray}
 \label{sootnosheniespleteniya} LH_0=H_1L,
  \end{eqnarray}
then
\begin{eqnarray}
H_1=-i\alpha _1 \partial_x+V_1,\quad
V_1=V_0+\left[-i\alpha_1,\frac{du(x)}{dx}u^{-1}(x)\right].
 \end{eqnarray}
The transformed spinor functions
\begin{eqnarray}
 \label{preobrazovannyespinornyefunkcii}\tilde{\psi}=L\psi
 \end{eqnarray}
are solutions of the transformed Dirac equation
\begin{eqnarray}\label{90}
H_1\tilde{\psi}(x)=E\tilde{\psi}(x).
\end{eqnarray}
 The Darboux transformations of the four--component Dirac equation is
similar to the Darboux transformations of the two--component Dirac
equation \cite{annphys2003v305p151}.

Consider the free Dirac Hamiltonian $H_0$, $V_0=m\beta,$ where $m$
is a mass of particle,
$\beta=\left(%
\begin{array}{cc}
  \mathbf{I} & 0 \\
  0 & -\mathbf{I} \\
\end{array}%
\right),$ $\mathbf{I}=\left(%
\begin{array}{cc}
  1 & 0 \\
  0 & 1 \\
\end{array}%
\right).$ Suppose that $\Lambda$ is
\begin{equation}\label{lambda}
    \Lambda(\lambda_1,\lambda_2)=[\lambda_1(I+\beta)+\lambda_2(I-\beta)]/2,
\end{equation}
where $\lambda_1=\pm\varepsilon_0,$ $\lambda_2=\pm\varepsilon_1,$
$\varepsilon_n=\sqrt{m^2-k^2n^2},$ $k$ is real number $(k\ll
m).$\\
Four different sets of parameters $\lambda_1,$ $\lambda_2$
correspond to four different matrices $\Lambda^{(i)}$
 ($i=\overline{1,4}$):
 \begin{eqnarray}
   \Lambda^{(1)}&=&\Lambda(\varepsilon_0,\varepsilon_1),\quad  \Lambda^{(2)}=\Lambda(\varepsilon_0,-\varepsilon_1), \\
   \Lambda^{(3)}&=&\Lambda(-\varepsilon_0,\varepsilon_1), \quad  \Lambda^{(4)}=\Lambda(-\varepsilon_0,-\varepsilon_1)
 \end{eqnarray}
and four corresponding transformation matrix functions $u^{(i)}$ are
solutions of the matrix Dirac equation
\begin{eqnarray}
  H_0u^{(i)}&=&\Lambda^{(i)}u^{(i)},\quad i=\overline{1,4}.
\end{eqnarray}
All these solutions have the common matrix structure
\begin{eqnarray}
  u^{(i)}&=&a^{(i)}I+b^{(i)}\alpha+c^{(i)}\beta+d^{(i)}\gamma,
\end{eqnarray}
where \begin{eqnarray}
  \alpha&=&\alpha_1,\quad \gamma=\alpha\beta,\quad
  \alpha^2=\beta^2=-\gamma^2=I.
\end{eqnarray}
It can be shown the inverse matrices have the following general
form:
\begin{eqnarray}
  {u^{(i)}}^{-1}&=&(a^{(i)}I-b^{(i)}\alpha-c^{(i)}\beta-d^{(i)}\gamma)/D,\\
  D&=&(a^{(i)2}-b^{(i)2}-c^{(i)2}+d^{(i)2}).\nonumber
\end{eqnarray}
where
\begin{eqnarray}
  a^{(i)}&=&\frac{1}{2}(\mu^{(i)}_1+\mu^{(i)}_3), \quad  b^{(i)}=\frac{1}{2}(\mu^{(i)}_2+\mu^{(i)}_4), \\
  c^{(i)}&=&\frac{1}{2}(\mu^{(i)}_1-\mu^{(i)}_3), \quad
  d^{(i)}=\frac{1}{2}(\mu^{(i)}_4-\mu^{(i)}_2).
\end{eqnarray}
Explicit values of magnitudes $\mu^{(i)}_k$
\begin{eqnarray}
  \mu^{(1)}_1&=&\mu^{(2)}_1=1=\mu^{(3)}_4=\mu^{(4)}_4,\quad
  \mu^{(1)}_4=\mu^{(2)}_4=0=\mu^{(3)}_1=\mu^{(4)}_1,\\
  \mu^{(1)}_3&=&\mu^{(2)}_3=\mu^{(3)}_3=\mu^{(4)}_3=\cosh{kx},
  \quad  \mu^{(1)}_2=\frac{ik\sinh{kx}}{\varepsilon_0-\varepsilon_1},\\
  \mu^{(2)}_2&=&\frac{ik\sinh{kx}}{\varepsilon_0+\varepsilon_1}, \quad
   \mu^{(3)}_2=-\frac{ik\sinh{kx}}{\varepsilon_0+\varepsilon_1},
   \quad  \mu^{(4)}_2=\frac{ik\sinh{kx}}{-\varepsilon_0+\varepsilon_1}.
\end{eqnarray}
Constructing the Darboux transformation operators with the help of
the obtained above matrices
\begin{equation}\label{operatorpreobrazovania}
    L^{(i)}=\frac{d}{dx}-\frac{du^{(i)}}{dx}(u^{(i)})^{-1},
\end{equation}
and the intertwining relations
\begin{equation}\label{cootnosheniaspletenia}
    L^{(i)}H^{(i)}_0=H^{(i)}_1L^{(i)}
\end{equation}
we generate  four new Hamiltonians
\begin{eqnarray}
  H^{(i)}_1&=&H^{(i)}_0+\left[-i\alpha,\frac{du^{(i)}}{dx}(u^{(i)})^{-1}\right],\quad
  i=\overline{1,4},
\end{eqnarray}
\begin{eqnarray}
    H_1^{(1)}&=&-i\alpha_1\partial_x-\varepsilon_1\beta+ik\tanh{(kx)}\gamma,\\
    H_1^{(2)}&=&-i\alpha_1\partial_x+\varepsilon_1\beta+ik\tanh{(kx)}\gamma,\\
    H_1^{(3)}&=&-i\alpha_1\partial_x-\varepsilon_1\beta-ik\tanh{(kx)}\gamma,\\
    H_1^{(4)}&=&-i\alpha_1\partial_x+\varepsilon_1\beta-ik\tanh{(kx)}\gamma.
\end{eqnarray}
The  Hamiltonians $H^{(1)}_1,$ $H^{(3)}_1,$  and $H^{(2)}_1,$
$H^{(4)}_1,$ are related by unitary transformation
\begin{eqnarray}
  \label{36}H_1^{(3)}=UH_1^{(1)}U^{-1},\quad
  H_1^{(4)}=UH_1^{(2)}U^{-1},\quad U=\alpha
\end{eqnarray}
and consequently they are isospectral.

Spectra of the  Hamiltonians   $H^{(1)}_1,$ $H^{(3)}_1,$
$H^{(2)}_1,$ $H^{(4)}_1,$ in difference of  $H^{(0)}$ contain both
two continuous branches $m\leq E<\infty,$ $-\infty<E\leq-m$ and
discrete part containing one level.\\ Energies of bind states of
the Hamiltonians $H_1^{(1)}$ and $H_1^{(3)}$ are equal
$\varepsilon_1.$ Energies of bind states of the Hamiltonians
$H^{(2)}_1,$ $H^{(4)}_1$ are equal $-\varepsilon_1.$

Besides of relations \eqref{36},  between Hamiltonian pairs
$H^{(i)}_1$ there are intertwining relations
\begin{eqnarray}
  L^{(1,4)}_1H^{(1)}_1&=&H^{(4)}L^{(1,4)}_1,\quad  L^{(2,3)}_1H^{(2)}_1=H^{(3)}L^{(2,3)}_1, \\
  L^{(4,1)}_1H^{(4)}_1&=&H^{(1)}_1L^{(4,1)}_1, \quad  L^{(3,2)}_1H^{(3)}_1=H^{(3)}_1L^{(3,2)}_1,
\end{eqnarray}
\begin{eqnarray}
  L^{(1,2)}_1H^{(1)}_1&=&H^{(2)}_1L^{(1,2)}_1,\quad  L^{(2,1)}_1H^{(2)}_1=H^{(1)}_1L^{(2,1)}_1, \\
  L^{(3,4)}_1H^{(3)}_1&=&H^{(4)}L^{(3,4)}_1, \quad  L^{(4,3)}_1H^{(4)}_1=H^{(3)}L^{(4,3)}_1.
\end{eqnarray}
Here
\begin{eqnarray}
  L^{(1,4)}_1&=&L^{(2,3)}_1=\bar{L}_1,\quad  L^{(4,1)}_1=L^{(3,2)}_1=\tilde{L}_1, \\
  L^{(1,2)}_1&=&\bar{L_1}U,\quad L^{(2,1)}_1=\tilde{L}_1U,\\
   L^{(3,4)}_1&=&\bar{L_1}U,\quad L^{(4,3)}_1=\tilde{L}_1U,\\
\bar{L}_1&=&\frac{d}{dx}-\frac{d\bar{u}}{dx}(\bar{u})^{-1},\quad
\tilde{L}_1=\frac{d}{dx}-\frac{d\tilde{u}}{dx}(\tilde{u})^{-1},
\end{eqnarray}
where
\begin{eqnarray}
  \bar{u}&=&\cosh{(kx)}(I+\beta)+\frac{1}{\cosh{(kx)}}(I-\beta), \\
  \tilde{u}&=&
  \cosh{(kx)}(\alpha+\gamma)+\frac{1}{\cosh{(kx)}}(\alpha-\gamma).
\end{eqnarray}
\section{Multi--soliton shape--invariant\\Hamiltonians}
Consider the following Hamiltonians $H^{(i)}_n$ $(i=\overline{1,4})$
\begin{eqnarray}
 \label{51} H^{(1)}_n&=&-i\alpha_1\frac{d}{dx}-\varepsilon_n\beta+ink\tanh{(kx)}\gamma,\\
  \label{52}H^{(2)}_n&=&-i\alpha_1\frac{d}{dx}+\varepsilon_n\beta+ink\tanh{(kx)}\gamma, \\
  H^{(3)}_n&=&-i\alpha_1\frac{d}{dx}+\varepsilon_n\beta-ink\tanh{(kx)}\gamma, \\
  H^{(4)}_n&=&-i\alpha_1\frac{d}{dx}-\varepsilon_n\beta-ink\tanh{(kx)}\gamma,
\end{eqnarray}
coinciding  with the obtained Hamiltonians $H^{(i)}_1$ at $n=1.$\\
It is evidently that
\begin{eqnarray}
  \label{55}H_n^{(3)}=UH_n^{(1)}U^{-1},\quad
  H_n^{(4)}=UH_n^{(2)}U^{-1},\quad U=\alpha_1.
\end{eqnarray}
It is easily to calculate that the intertwining relations
 $ L^{(i,k)}_nH^{(i)}_n=H^{(k)}_{n+1}L^{(i,k)}_n$ ($i,k=1,2$ or
 $i,k=3,4$) have place.

At that \begin{eqnarray}\label{spletenie2}
  \bar{L}_nH^{(1)}_n&=&H^{(4)}_{n}\bar{L}_n,\quad \tilde{L}_nH^{(4)}_n=H^{(1)}_{n}\tilde{L}_n,\\
  \bar{L}_nH^{(2)}_n&=&H^{(3)}_{n}\bar{L}_n,\quad
  \tilde{L}_nH^{(3)}_n=H^{(2)}_{n}\tilde{L}_n\nonumber,
\end{eqnarray}
where
\begin{eqnarray}
\bar{L}_n&=&\frac{d}{dx}-\frac{d\bar{u}_n}{dx}(\bar{u}_n)^{-1},\quad\tilde{L}_n=\frac{d}{dx}-\frac{d\tilde{u}_n}{dx}(\tilde{u}_n)^{-1},
\end{eqnarray}
\begin{eqnarray}
  \bar{u}_n&=&\cosh^n{(kx)}(I+\beta)+\coth^{-n}{(kx)}(I-\beta), \\
  \tilde{u}_n&=&
  \cosh^n{(kx)}(\alpha+\gamma)+\coth^{-n}{(kx)}(\alpha-\gamma),
\end{eqnarray}
\begin{eqnarray}
  L^{(i,k)}_n&=&\frac{d}{dx}-\frac{d}{dx}(u^{(i,k)}_n)(u^{(i,k)}_n)^{-1},\\
    u^{(i,k)}_n&=&a^{(i,k)}_nI+b^{(i,k)}_n\alpha+c^{(i,k)}_n\beta+d^{(i,k)}_n\gamma,\\
  a^{(i,k)}_n&=&\frac{1}{2}(\mu^{(i,k)}_{n,1}+\mu^{(i,k)}_{n,3}),\quad
  b^{(i,k)}_n=\frac{1}{2}(\mu^{(i,k)}_{n,2}+\mu^{(i,k)}_{n,4}), \\
  c^{(i,k)}_n&=&\frac{1}{2}(\mu^{(i,k)}_{n,1}-\mu^{(i,k)}_{n,3}),\quad
  d^{(i,k)}_n=\frac{1}{2}(\mu^{(i,k)}_{n,4}-\mu^{(i,k)}_{n,2}).
\end{eqnarray}
The obvious expressions for quantities  $\mu^{(i,k)}_{n,j}$
$(j=\overline{1,4})$ are presented in Appendix.

The transformation matrixes $u^{(i,k)}_{n}$ are  solutions of the
matrix Dirac equations
\begin{eqnarray}
  H^{(i)}_nu^{(i,k)}_{n}&=&u^{(i,k)}_{n}\Lambda^{(i,k)}_n,\quad
   \Lambda^{(i,k)}_n=\frac{1}{2}\left[(I+\beta)\lambda^{(i,k)}_{n,1}+(I-\beta)\lambda^{(i,k)}_{n,2}\right].
\end{eqnarray}
The values of quantities $\lambda^{(i,k)}_{n,j}$ $(j=1,2)$ are in
Appendix.

Intertwining relations \eqref{spletenie2} are presented in the
diagram form
\begin{eqnarray}
\nonumber&\stackrel{L^{(2)}_0}\nearrow
H^{(2)}_1\stackrel{L^{(2)}_1}\longrightarrow
H^{(2)}_2\stackrel{L^{(2)}_2}\longrightarrow
\ldots\stackrel{L^{(2)}_{n-1}}\longrightarrow
H^{(2)}_n\stackrel{L^{(2)}_{n}}\longrightarrow
H^{(2)}_{n+1}\ldots\\\nonumber
H_{0}&{L^{(1,2)}_1}\nearrow\quad\quad\searrow
~_{L^{(2,1)}_2}\quad\quad\quad _{L^{(1,2)}_n}\nearrow\\\nonumber
&\stackrel{L^{(1)}_0}\searrow
H^{(1)}_1\stackrel{L^{(1)}_1}\longrightarrow
 H^{(1)}_2\stackrel{L^{(1)}_2}\longrightarrow\ldots\stackrel{L^{(1)}_{n-1}}\longrightarrow H^{(1)}_n\stackrel{L^{(1)}_n}
 \longrightarrow H^{(1)}_{n+1}\ldots\\\nonumber
  &^{\widetilde{L}_2}\uparrow\downarrow_{\bar{L}_2}\qquad\quad\quad\quad~^{\tilde{L}_n}
  \uparrow\downarrow_{\bar{L}_{n}}~\\\nonumber
  &\stackrel{L^{(4)}_0}\nearrow H^{(4)}_1\stackrel{L^{(4)}_1}\longrightarrow
  H^{(4)}_2\stackrel{L^{(4)}_2}\longrightarrow \ldots\stackrel{L^{(4)}_{n-1}}\longrightarrow H^{(4)}_n\stackrel{L^{(4)}_n}\longrightarrow H^{(4)}_{n+1}\ldots\\\nonumber
  H_{0}&_{L^{(3)}_1}\nearrow\quad\qquad\searrow
~_{L^{(3)}_2}\quad\quad\quad _{L^{-}_n}\nearrow\\\nonumber
  &\stackrel{L^{(3,4)}_0}\searrow H^{(4,3)}_1\stackrel{L^{(3,4)}_1}\longrightarrow
  H^{(3)}_2\stackrel{L^{(3)}_2}\longrightarrow\ldots\stackrel{L^{(3)}_{n-1}}\longrightarrow H^{(3)}_n\stackrel{L^{(3)}_n}\longrightarrow H^{(3)}_{n+1}\ldots\nonumber
\end{eqnarray}
 Hence with the help of the Darboux transformation technique (the Darboux--Crum chains \cite{Crum})
we construct two-parameter family of the Dirac Hamiltonians
$H^{(i)}_{n},$ $i=\overline{1,4},$ which are $(2n-1)$-solitons, i.
e. they have $2n-1$ bound states.
\section{Spectral structure and \\eigenfunctions of Hamiltonians $H^{(i)}_n$}
To investigate the solutions of the equations
\begin{eqnarray}
\label{*}
  H^{(i)}_n\psi^{(i)}&=&E_n\psi^{(i)},\quad  i=1,2,
\end{eqnarray}
we have
\begin{eqnarray}
  \psi^{(3)}&=&\alpha\psi^{(1)},\quad \psi^{(4)}=\alpha\psi^{(2)}.
\end{eqnarray}

From \eqref{*} and \eqref{51}, \eqref{52} it is easily to obtain
that the components of spinors
$$\psi^{(i)}=(\psi^{(i)}_1,\psi^{(i)}_2,\psi^{(i)}_3,\psi^{(i)}_4)^T$$
at $i=1,2$  are the solutions of second-order equations
\begin{eqnarray}
  (\psi^{(i)}_j)^{\prime\prime}+\frac{n(n-1)k^2}{\cosh^2{kx}}\psi^{(i)}_j+(E^2-m^2)\psi^{(i)}_j&=&0,\quad j=1,2, \\
  (\psi^{(i)}_j)^{\prime\prime}+\frac{n(n+1)k^2}{\cosh^2{kx}}\psi^{(i)}_j+(E^2-m^2)\psi^{(i)}_j&=&0,\quad j=3,4,
\end{eqnarray}
which  with the help of substitution $t=\tanh{kx}$ are reduced to
the
equations for associated Legendre polynomials.\\
 The energy spectrum at $|E|\geq m$  is continuous and the energy spectrum at $|E|<m$
 is discrete.\\
The energy levels are
\begin{equation}\label{energii}
    E_s=\sign(s)\varepsilon_s=\sign(s)\sqrt{m^2-k^2s^2},
\end{equation}
where
\begin{eqnarray}
  s=\left\{\begin{array}{cc}
     -n+1,-n+2,\ldots,-1,1,\ldots,n-1,n,  \quad i=1,\\
     -n,-n+1,\ldots,-1,1,\ldots,n-2,n-1,  \quad i=2.\\
  \end{array}\right.
\end{eqnarray}
The solutions of equations
\begin{eqnarray}\label{H1}
H^{(1)}_n\psi_s=E_s\psi^{(1)}_s
\end{eqnarray}
are
\begin{eqnarray}
  \psi^{(1)}_{s}&=&(\psi^{(1)}_{s1},\psi^{(1)}_{s2},\psi^{(1)}_{s3},\psi^{(1)}_{s4})^T,\\
  \psi^{(1)}_{s1}&=&A^{(1)}_sP^s_{n-1}(t), \quad  \psi^{(1)}_{s2}=B^{(1)}_sP^s_{n-1}(t), \\
  \psi^{(1)}_{s3}&=&C^{(1)}_sP^s_{n}(t), \quad \psi^{(1)}_{s4}=D^{(1)}_sP^s_{n}(t),\quad t=\tanh{kx},\\
  A^{(1)}_s&=&\frac{ik(n+s)}{E_s+\varepsilon_n}D^{(1)}_s,\quad
  B^{(1)}_s=\frac{ik(n+s)}{E_s+\varepsilon_n}C^{(1)}_s.
\end{eqnarray}
The solutions of equations
\begin{eqnarray}\label{H1}
H^{(2)}_n\psi_s=E_s\psi^{(2)}_s
\end{eqnarray}
are
\begin{eqnarray}
  \psi^{(2)}_{s}&=&(\psi^{(2)}_{s1},\psi^{(2)}_{s2},\psi^{(2)}_{s3},\psi^{(2)}_{s4})^T,\\
  \psi^{(2)}_{s1}&=&A^{(2)}_sP^s_{n-1}(t), \quad \psi^{(2)}_{s2}=B^{(2)}_sP^s_{n-1}(t), \\
  \psi^{(2)}_{s3}&=&C^{(2)}_sP^s_{n}(t), \quad \psi^{(2)}_{s4}=D^{(2)}_sP^s_{n}(t),\quad t=\tanh{kx},\\
  A^{(2)}_s&=&\frac{ik(n-s)}{E_s-\varepsilon_n}D^{(2)}_s,\quad
  B^{(2)}_s=\frac{ik(n-s)}{E_s-\varepsilon_n}C^{(2)}_s.
\end{eqnarray}
Here $P^s_{n-1}(t)$ and $P^s_{n}(t)$ are associated Legendre
polynomials \cite{GRAND}.\\
Continuous spectrum is doubly degenerate,  i. e.  for each value $E$
($|E|\geq m$) equations \eqref{*} have two linear-independent
solutions, that can be obtained from the wave functions of discrete
spectrum by substitution
\begin{eqnarray}
  E_s\rightarrow E,\quad s\rightarrow\pm i\mu,\quad P^s_n\rightarrow
  P^{\pm i\mu}_n(t),
\end{eqnarray}
\begin{eqnarray}
  P^{\pm i\mu}_n(t)&=&\frac{1}{\Gamma(1\mp i\mu)}\exp{(\pm ipx)}F(-n,n+1;1\mp
  i\mu;(1-t)/2).
\end{eqnarray}
From the last expression it is evident that the obtained potentials
 are transparent.
\section{Conclusion}
In the paper we constructed the special Darboux transformation
chains. These chains generate the set of the shape invariant
exactly solvable four-compo\-nent Dirac Hamiltonians. The
Hamiltonians may be applied  for investigation of spin $1/2$
relativistic particle moving in one-dimensional periodic
structures \cite{Samsonov,Eurjphys,bagrovpecer}. The results
corresponding investigations will be published.

In conclusion the author is grateful to Professors A. Tarasov for
fruitful discussions,  V.G. Bagrov for critical remarks, O.O.
Voskresenskaya and S.R. Gevorkyan for helpful discussion. I am
grateful to the Joint Institute for Nuclear Research (Dubna, Moscow
region, Russia) for hospitality during this work. The work was
supported in part by the ‘Dynasty’ Fund and Moscow International
Center of Fundamental Physics.
\section*{Appendix}
In this Appendix we demonstrate the explicit forms
\begin{enumerate}
    \item [(i)] for quantities $\lambda^{(i,k)}_{n,j}:$
\begin{eqnarray}
  \lambda^{(1,1)}_{n,1}&=&\lambda^{(1,2)}_{n,1}=\lambda^{(3,3)}_{n,1}=\lambda^{(3,4)}_{n,1}=-\varepsilon_n, \\
 \lambda^{(2,2)}_{n,1}&=&\lambda^{(2,1)}_{n,1}=\lambda^{(4,4)}_{n,1}=\lambda^{(4,3)}_{n,1}=\varepsilon_n, \\
  \lambda^{(1,1)}_{n,2}&=&\lambda^{(2,1)}_{n,2}=\lambda^{(3,3)}_{n,2}=\lambda^{(4,3)}_{n,2}=\varepsilon_{n+1}, \\
  \lambda^{(2,2)}_{n,2}&=&\lambda^{(1,2)}_{n,2}=\lambda^{(4,4)}_{n,2}=\lambda^{(3,4)}_{n,2}=-\varepsilon_{n+1},
\end{eqnarray}
    \item [(ii)] and also for quantities $\mu^{(i,k)}_{n,j}$ $(i,k=1,2):$
\begin{eqnarray}&&\mu^{(i,k)}_{n,1}=\cosh^n{kx},\quad\mu^{(i,k)}_{n,2}=\frac{(2n+1)ik
\sinh{kx}\cosh^n{kx}}{\lambda^{(i,k)}_{n,1}-\lambda^{(i,k)}_{n,2}},\\
&&\mu^{(i,k)}_{n,3}=\cosh^{n+1}{kx},\quad\mu^{(i,k)}_{n,4}=0,\end{eqnarray}
    \item [(iii)] and  $\mu^{(i,k)}_{n,j}$ $(i,k=3,4):$
\begin{eqnarray}&&\mu^{(i,k)}_{n,1}=0,\quad\mu^{(i,k)}_{n,2}=\cosh^{n+1}{kx},\\
&&\mu^{(i,k)}_{n,3}=\frac{(2n+1)ik\sinh{kx}\cosh^n{kx}}{\lambda^{(i,k)}_{n,1}-\lambda^{(i,k)}_{n,2}},\quad
\mu^{(i,k)}_{n,4}=\cosh^{n}{kx}.\end{eqnarray}
\end{enumerate}

\end{document}